\documentclass[twocolumn,prd,letterpaper,unsortedaddress,superscriptaddress]{revtex4}

\usepackage{graphicx}
\usepackage{dcolumn}
\usepackage{amsmath}
\usepackage{amssymb}
\usepackage{pslatex}

\def\deg{$^\circ$}
\newcommand{\sci}[2]{#1\times 10^{#2}}

\newcommand{\vvec}{\mathbf{v}}

\newcommand{\beq}{\begin{equation}}
\newcommand{\eeq}{\end{equation}}

\begin{document}

\title{Observations of Microwave Continuum Emission from Air Shower Plasmas}

\author{P.~W.~Gorham}
\affiliation{University of Hawaii at Manoa, 
Department of Physics and Astronomy,
Honolulu, Hawaii 96822}

\author{N.~G.~Lehtinen}
\altaffiliation{currently at Stanford University}
\affiliation{University of Hawaii at Manoa, 
Department of Physics and Astronomy,
Honolulu, Hawaii 96822}

\author{G.~S.~Varner}
\affiliation{University of Hawaii at Manoa, 
Department of Physics and Astronomy,
Honolulu, Hawaii 96822}

\author{J.~J.~Beatty}
\affiliation{Ohio State University, Dept. of Physics, Columbus, OH.}

\author{A.~Connolly}
\affiliation{Univ. of Calif. at Los Angeles, Dept. of Physics,
Los Angeles, CA}

\author{P.~Chen}
\affiliation{Stanford Linear Accelerator Center,
2575 Sand Island Rd., Menlo Park, CA.}

\author{M.~E.~Conde}
\affiliation{Argonne National Laboratory,
Argonne, IL.}

%\author{R.~C.~Field}
%\affiliation{Stanford Linear Accelerator Center,
%2575 Sand Island Rd., Menlo Park, CA.}

\author{W.~Gai}
\affiliation{Argonne National Laboratory,
Argonne, IL.}

\author{C.~Hast}
\affiliation{Stanford Linear Accelerator Center,
2575 Sand Island Rd., Menlo Park, CA.}

\author{C.~L.~Hebert}
\altaffiliation{currently with Raytheon Corp., Tucson, AZ}
\affiliation{University of Hawaii at Manoa, 
Department of Physics and Astronomy,
Honolulu, Hawaii 96822}

\author{C.~Miki}
\affiliation{University of Hawaii at Manoa, 
Department of Physics and Astronomy,
Honolulu, Hawaii 96822}

\author{R.~Konecny}
\affiliation{Argonne National Laboratory,
Argonne, IL.}

\author{J.~Kowalski}
\affiliation{University of Hawaii at Manoa, 
Department of Physics and Astronomy,
Honolulu, Hawaii 96822}

\author{J.~Ng}
\affiliation{Stanford Linear Accelerator Center,
2575 Sand Island Rd., Menlo Park, CA.}

\author{J.~G.~Power}
\affiliation{Argonne National Laboratory,
Argonne, IL.}

\author{K.~Reil}
\affiliation{Stanford Linear Accelerator Center,
2575 Sand Island Rd., Menlo Park, CA.}

\author{D.~Saltzberg}
\affiliation{Univ. of Calif. at Los Angeles, Dept. of Physics,
Los Angeles, CA}

\author{B.~T.~Stokes}
\altaffiliation{currently at John A. Burns School of Medicine,
UH Manoa}
\affiliation{University of Hawaii at Manoa, 
Department of Physics and Astronomy,
Honolulu, Hawaii 96822}

\author{D.~Walz}
\affiliation{Stanford Linear Accelerator Center,
2575 Sand Island Rd., Menlo Park, CA.}

\begin{abstract}
We investigate a possible new technique for microwave detection of
cosmic ray extensive air showers which relies on detection of
expected continuum radiation in the microwave range, caused by
free-electron collisions with neutrals in the tenuous plasma
left after the passage of the shower. 
We performed an initial
experiment at the AWA (Argonne Wakefield Accelerator) laboratory
in 2003 and measured broadband microwave emission from air ionized
via high energy electrons and photons.  
A follow-up experiment at SLAC (Stanford Linear Accelerator
Center) in summer of 2004 confirmed the major features of the previous 
AWA observations with better precision. Prompted by these results
we built a prototype detector using satellite television 
technology, and have made measurements suggestive of detection
of cosmic ray extensive air showers. The method, if confirmed
by experiments now in progress, could provide a high-duty cycle
complement to current nitrogen fluorescence observations.
\end{abstract}

\maketitle

\section{Introduction}
The origin and nature of the ultra-high energy cosmic rays (UHECR)
remains one of the enduring mysteries of experimental particle
astrophysics. In spite of well over four decades of observations 
of $10^{20}$~eV UHECR by many different experiments~\cite{Linsley_62},
we still do not have a confirmed astrophysical source for
these particles, nor do we understand their composition in any
detail, nor do we know how they propagate from their unknown sources
to earth~\cite{Cronin:2004ye}. In the last decade new observatories
such as HiRes and most recently the Auger Observatory have 
much improved the statistics on measurements of these particles,
but the issues of their origin and propagation remain largely open.
As the highest energy subatomic particles observed in nature, UHECRs
must arise from the most energetic phenomena in our universe.
Their study is thus crucial to understanding the nature of 
acceleration processes that can attain energies some seven orders of
magnitude higher than is currently achievable 
in the laboratory~\cite{Chung:1997rz,Stanev:1996qj,Stecker:2001vb}.  

The primary UHECR spectrum
is described by a simple power law
$J(E) \propto E^{-\alpha}$
with $\alpha\simeq2.7$ for $10^{18.5}<E<10^{19.5}$~eV 
\cite{Abbasi:2005bw,Takeda:1998ps}. Above $10^{19.5}$,
the interaction length of cosmic ray nucleons on the
cosmic microwave background becomes comparable to
intergalactic separation distances, a process
first described by Greisen~\cite{Greisen:1966jv}, 
and Zatsepin \& Kuzmin~\cite{Zatsepin:1966jv} and now known as
the GZK process. It is precisely at and above the GZK
energies that the measurements of the primary UHECR become
uncertain due to low statistics, and the shape of
the spectrum near the endpoint is still a subject of
active debate. 

\begin{figure}[tb!]
  \includegraphics[width=0.9\columnwidth]{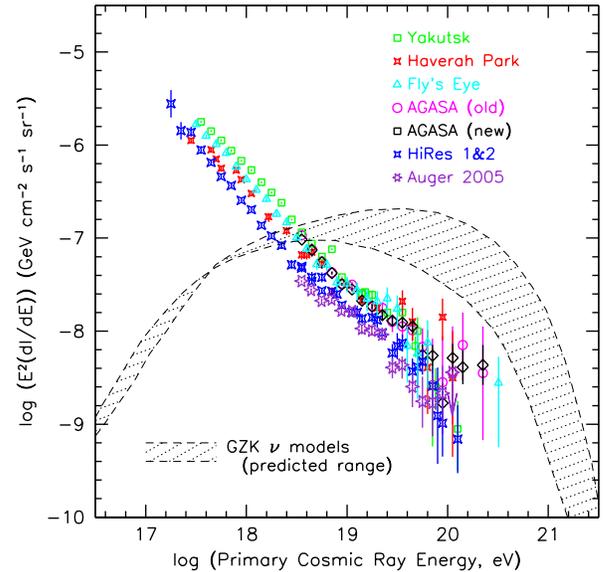}
\begin{small}
\caption{\it 
World ultra-high energy cosmic ray and
predicted cosmogenic neutrino spectrum as of early 2007,
including data from the Yakutsk~\cite{Yakutsk},
Haverah Park~\cite{Haverah} 
the Fly's Eye~\cite{Flyseye}, AGASA~\cite{Abbasi:2002ta},
HiRes~\cite{Abbasi:2005bw}, and Auger~\cite{Sommers:2005vs}, 
collaborations. Data points represent differential 
flux $dI(E)/dE$, multiplied by $E^{2}$. Error bars are statistical only.
GZK neutrino models are 
from Protheroe \& Johnson~\cite{Proth_gzk} and Kalashev et al.~\cite{Kal02}.
    \label{fig:spec1}
  }
\end{small}
\end{figure}

Because of the scarcity of particles 
at these highest energies, research into new methods
has focused on indirect means of
observation \cite{Corbato:1992fq,Cronin:1995zs},
which makes use of radiated air-fluorescence emission from the 
air shower to observe it at distances of up to tens of
km from the particle axis. 
By observing the longitudinal and transverse development of 
UHECR-induced extensive air showers (EASs)  
investigators are gaining information on the primary
composition, which favors light elements and disfavors
a significant electromagnetic (e.g. photon) component. 
Such studies can also elucidate the
high energy physics of the early interactions, which
occur at center-of-mass energies well above that
currently probed by accelerators\cite{Cronin:2004ye}.

The region near the endpoint of the UHECR
energy spectrum is shown in summary form in Fig.~\ref{fig:spec1},
where no effort has been made to correct the systematic offsets
in the flux levels of the different experiments involved.
Above $10^{20}$~eV, the 
event rate is of order 1 per km$^2$ per century, producing still
only a handful of events per year close to this threshold in
all existing UHECR observatories. As is apparent from the current
world spectrum, constraints on the high energy tail or 
statistically compelling details of any putative cutoff above the current 
highest energies will still require years of observation.
The need for much-improved statistics to address the primary issues
currently under investigation all argue for expansions of and
improvements on existing methods. Such issues include the
detailed shape of the UHECR energy spectrum
\cite{Takeda:1998ps,Abbasi:2002ta}
(including the presence, or lack thereof, of the
GZK cutoff~\cite{Greisen:1966jv,Zatsepin:1966jv}),
energy-resolved primary particle composition~\cite{Abbasi:2004nz,Abu-Zayyad:2000ay},
and source production mechanisms 
(i.e. origins)~\cite{Abbasi:2004vu,Abbasi:2004ib,Abbasi:2003tk,Uchihori:1999gu}.

In addition, the virtual certainty of the extragalactic
origin of these particles ensures an associated cosmogenic
neutrino flux, generated via photohadronic processes throughout
the universe~\cite{Engel_01}. Hadronic cosmic rays above $\sim 10^{19}$~eV 
propagating in the
2.7K cosmic microwave background radiation (CMBR) exceed
the threshold for resonant $\Delta^+$ particle production
through the GZK process,
and the rapid decay of these unstable secondaries leads to 
pions and subsequently neutrinos. The mean free path of
a $10^{20}$~eV proton in the CMBR is several Mpc in the
current epoch, whereas the neutrinos are unattenuated
from any cosmic distance. Future observations of cosmogenic GZK neutrinos
will provide a unique and complementary view of the UHECR
production, propagation, and attenuation throughout the
universe, motivated by the UHECR observations themselves.
Figure~\ref{fig:spec1} shows also a band indicating the
range of models for these cosmogenic neutrinos. The
uncertainties in the models stem primarily from the
details of the highest energy part of the UHECR spectrum,
as well as the epoch of maximum UHECR source evolution,
and GZK neutrino observations will thus provide independent
constraints on the UHECR sources.

\subsubsection{Motivation for Microwave EAS Detection.}

While there is general agreement among the different experiments
as to the global properties of the UHECR spectrum, as Figure~\ref{fig:spec1}
shows, there is still significant disagreement and uncertainty on absolute
flux scales and on some of the fundamental questions of UHECR research.  
The two primary techniques 
of UHECR observation, ground-based particle
arrays and optical fluorescence detectors both suffer from tangible
limitations.  In the case of ground arrays, only a single slice of EAS 
longitudinal development can be observed. This means that 
determinations of primary particle energy and composition require extrapolation
via model-dependent estimates, which may disagree depending on the
model used. While the optical fluorescence method
enables one to observe longitudinal as well as transverse shower development,
it is highly constrained by the fact that it can only work on clear, moonless 
nights. This leads to a net yearly duty cycle of only 5-10\%\cite{Corbato:1992fq}.
Furthermore, because the highest energy events are observed at
increasingly large distances, even small fluctuations in atmospheric aerosol
contamination can have substantial effects on energy estimation.

An air shower dissipates virtually its entire energy budget through
ionization, producing a tenuous plasma with an
electron temperature of order $10^{5}$~K or more. The ionization 
and subsequent de-excitation of molecular nitrogen in the
N$_2^{+*}$ 1N and 2P states leads directly to the optical
N$_2$ fluorescence now observed. The hot air shower plasma
cools rapidly on 1-10 nanosecond time scales, distributing its thermal
energy through collisions with the neutral molecules, primarily 
N$_2$, which has the largest cross section and number density.
This rapid cooling process leads to 
additional excitation of rotational, vibrational,
electronic valence, and other modes of kinetic energy distribution among molecules,
many of which can also lead to rapid de-excitation and subsequent emission.  

In turn, the hot electrons themselves,
while producing this excitation, can produce their own emission,
such as continuum bremsstrahlung emission, or recombination radiation.
The fraction of total radiated energy in optical fluorescence, compared
to the total available energy budget for secondary radiation, is
very small, leaving much possible radiative energy still unaccounted for.
The possibilities for observing secondary air shower plasma emission other than
optical fluorescence have not yet been explored in any detail.

We report here on investigations of the feasibility of 
other channels for EAS
observations. To this end we have undertaken 
several experimental efforts, including
two accelerator experiments. The promising results 
from these measurements have led us to commission a testbed 
prototype detector, which has helped to establish the methodology that 
could be used  to make detailed measurements of EAS microwave molecular
bremsstrahlung radiation (MBR)\cite{DEC}.  In this report
we describe the accelerator results and the testbed development that
has resulted from them, which we have dubbed the 
Air-shower Microwave Bremsstrahlung Experimental Radiometer (AMBER).

\subsection{Molecular Bremsstrahlung Radiation.}
\begin{figure}[bth!]
  \includegraphics[width=3.2in]{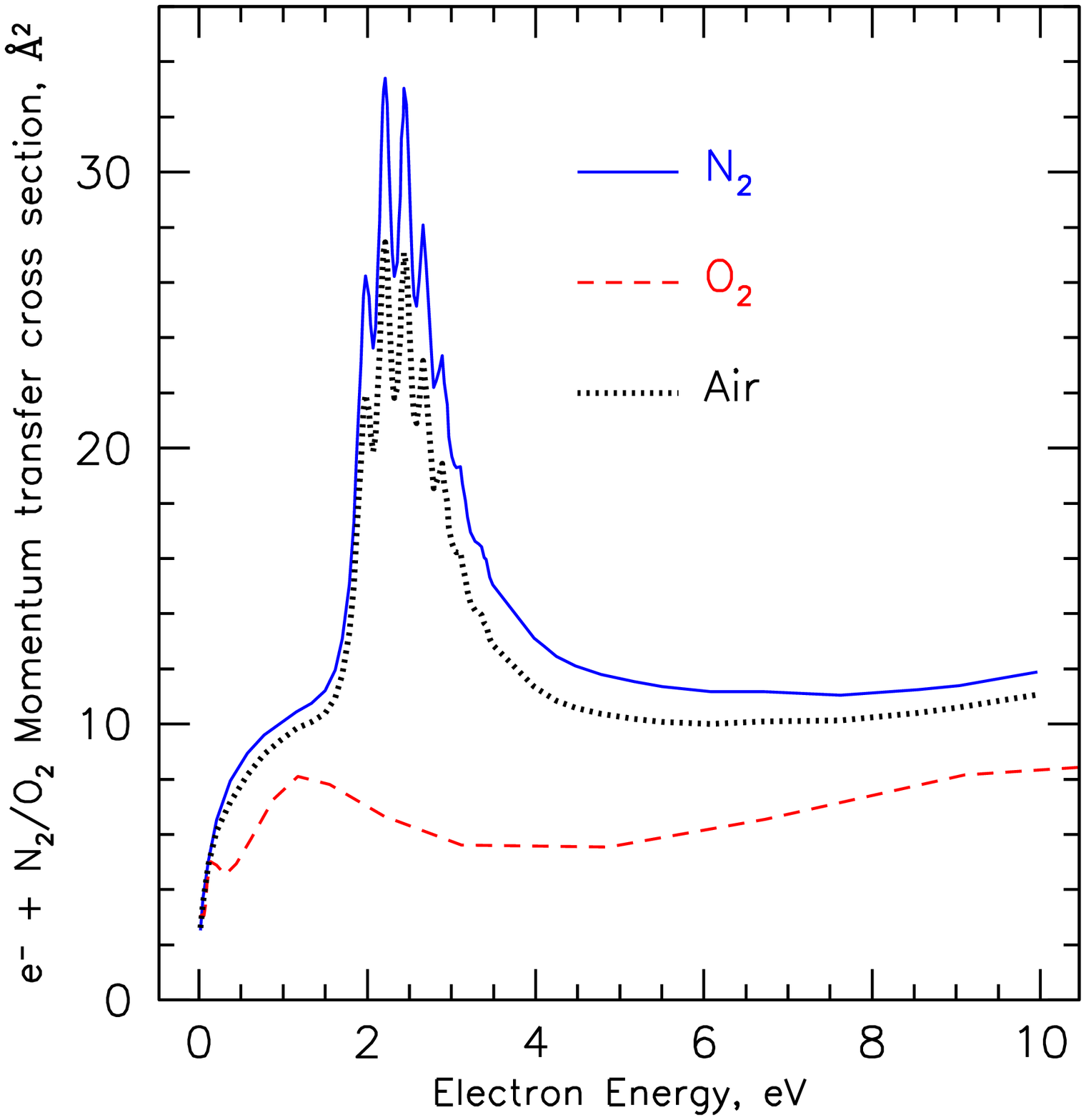}
  \vspace{2mm}
  \includegraphics[width=3.2in]{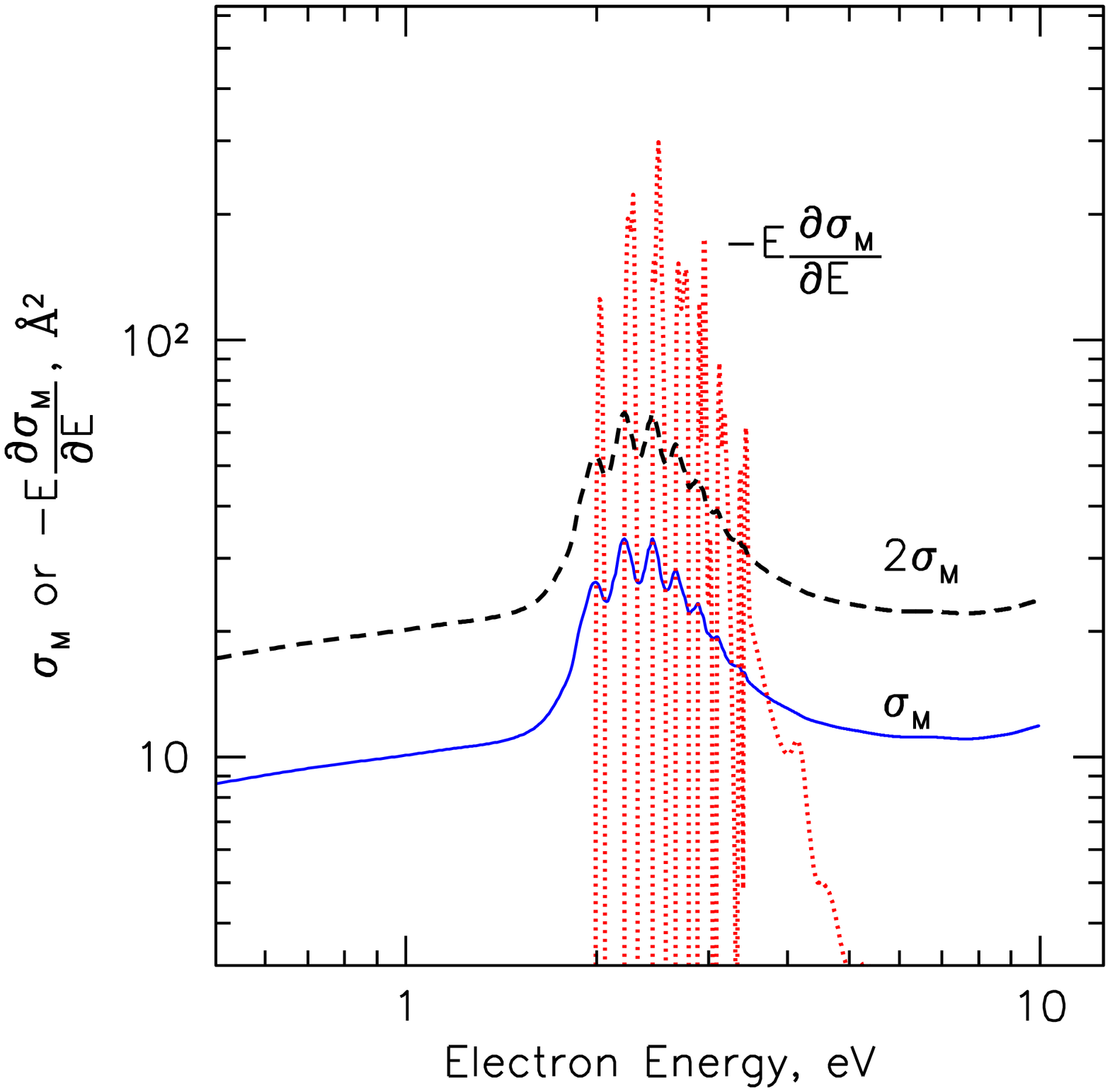}
  \caption{\it Top: Cross section for momentum transfer between
  electrons and N$_2$, O$_2$, and air over the energy range
  of interest for molecular bremsstrahlung production~\cite{N2x,O2x}.
  Bottom: The red-dotted lines show the function $-E \frac{\partial \sigma_M}{ \partial E}$
  which must be greater than either $\sigma_M$ (weak criterion; blue line) or
  $2\sigma_M$ (strong criterion, black-dash line) as a necessary condition for stimulated 
  bremsstrahlung emission in an air plasma. See text for details.\label{n2x}}
  
\end{figure}

MBR in weakly ionized air is created by free electrons 
accelerating through collisions with the fields of molecules in the ambient medium. 
EAS ionization is considered ``weak'' since the interactions of free electrons
or ions are dominated by collisions with neutrals rather than other ions.
MBR has been classically treated as a thermal process,
with the emission coming from $\leq 10$~eV electrons assumed to be
distributed with isotropic Maxwellian velocities. By these assumptions 
steady-state MBR emission is
expected to be isotropic and unpolarized, which strongly
differentiates it with the highly directional bremsstrahlung from relativistic particles 
which may be more familiar to a high energy or cosmic-ray physicist.
MBR emission shares this property with optical fluorescence emission,
an important feature which allows for the possibility of 
performing shower calorimetry by mapping the MBR intensity 
(and thus the ionization content) along a shower,
much as optical fluorescence detections maps the nitrogen excitation
along the same shower. As indicated in Fig.~\ref{Ambergeom}, MBR
emission and optical fluorescence both are emitted in all directions
around an EAS, and detectors may therefore ``image'' the air shower glow
as a track along the sky to establish two-dimensional angular geometry, 
and use the timing information for the pulse arrival to determine the
range evolution of the shower, giving a complete reconstruction of
both geometry and particle number evolution.

The expected isotropic behavior of MBR is also in sharp contrast to {\em relativistic}
radio emission processes such as Cherenkov, transition, or synchrotron radiation,
all of which are beamed and highly polarized. In this respect it
is convenient to think of MBR emission as analogous to ``radio fluorescence,''
whereas beamed relativistic radio emission is closely aligned to the particle
content of a shower and thus should be identified as concordant with
the information derived from a ground EAS detector array.
Furthermore, MBR intensity is expected to be proportional to 
the EAS ionization rate, which is known to be itself proportional to $N$, the
total number of charged particles in the shower. 
This therefore leads to a direct relationship between
MBR intensity and shower energy, with the degree of proportionality
determined by the details of local correlations between
electron velocities or radiative transitions in the tenuous air shower plasma.

The proportionality will depend on
important details which require empirical determination, much the same way
that oscillator line strengths necessary to understanding optical fluorescence
must be determined via laboratory calibration with additional 
corrections for atmospheric conditions. 
For example, since an EAS produces an initial
distribution of ionization which is likely to be a power-law rather than a
Maxwellian, there are corrections for non-thermal effects such
as stimulated emission and other non-equilibrium continuum radio emission channels,
which may significantly increase the emission power over the minimal thermal
MBR baseline. In addition, the cross sections for both elastic and inelastic
collisions of electrons with air molecules are complex functions of electron
energy, yielding strong velocity dependence in the electron collision frequency
which can also contribute to MBR emission coefficients.
Such effects are difficult to analytically estimate and will
be best calibrated {\em in situ} as has been done for other EAS observation
methods.

To analytically determine the expected minimum flux density for MBR, we start with
the  emissivity from classical bremsstrahlung analysis of
collisions between electrons and neutral molecules~\cite{DEC}:
\begin{equation}
\eta_\omega({\rm u})=\frac{e^2}{16\pi^3\varepsilon_0c^3}{\rm u}^2
\nu_{en}({\rm u})\zeta(\nu_{en},\omega)~~.
\label{eqn:nw1}
\end{equation}
Where $\omega$ is the microwave radian frequency,
${\rm u}$ is the electron velocity, $\nu_{en}({\rm u})$ is the velocity-dependent
electron-neutral collision frequency, 
and $\zeta(\nu_{en},\omega)$ is a term that accounts for
the collisional suppression of radiation
which arises from the destructive interference of fields from
successive collisions within the radiation formation
zone of each collision, a process also known as
{\em plasma dispersion}~\cite{DEC}.

Under the assumption of
an isotropic and time-stationary velocity distribution,
\begin{equation}
\zeta(\nu_{en},\omega) =  \frac{1} {1 + (\nu_{en}({\rm u})/\omega)^2}~~.
\label{zeta}
\end{equation}
For an altitude of 5~km, $\nu_{en}\simeq 3$~THz at electron energies of about
2~eV, near the peak of the collision cross section; for room temperature electrons,
$\nu_{en}\simeq 40$~GHz. The corresponding suppression factors are $\zeta\simeq 5 \times 10^{-5}$,
and $\zeta\simeq 0.4$, respectively, showing the wide extremes of values 
possible under the range of electron temperatures that obtain in
an air shower.

To preface further discussion below, we note that
there are several other effects that compete with plasma dispersion and will tend
to enhance the emissivity, or ``suppress the suppression.'' 
First, stimulated emission, even at a very small level, leads to correlations in
electron-photon transitions. Such emission does not require a full-blown
population inversion, as in laser processes. Rather, deviations from
the ground-state Maxwellian distribution can enable low levels
of stimulated emission.
Second, velocity correlations of the electrons due to the imposed geometry
of the shower tracks and the anisotropic distribution of ions can
impose some cylindrical symmetry in the distributions, in tension
with the assumption of uncorrelated isotropy in the electron
behavior.
Finally, weak plasma correlations on the scale of the Debye length can
lead to coherent motion of electrons over very small scales, but large
enough to overcome some fraction of the suppression effects.

To introduce an {\em ad hoc} accounting for the sum of all such effects,
we impose a {\em coherence factor} $\xi$ which modifies the $\zeta$ term:
\begin{equation}
\zeta_c(\nu_{en},\omega,\xi) =  \frac{\xi} {1 + (\nu_{en}({\rm u})/\omega)^2}~~.
\label{zetac}
\end{equation}
where $\xi>1$ parameterizes the level of excess emission
above the ``suppression floor'' determined by $\zeta$ in the absence of
any correlation of either electrons or electron-photon transitions.
The term $\xi$ can then be determined empirically to set the scaling of
the emission, with $\xi=1$ representing the MBR floor level.

The emissivity $\eta_{\omega}$ above must be integrated over the
distribution of electron velocities to yield the emission coefficient
$j_{\omega}$ (W m$^{-3}$ (radian/s)$^{-1}$ steradian$^{-1}$):
\begin{equation}
j_\omega = \int_0^{\infty} \eta_\omega({\rm u})f({\rm u}){\rm u}^2~d{\rm u},
\label{eqn:jw}
\end{equation}
where $f({\rm u})$ is the electron distribution function,
which is Maxwellian in the thermal limit:
\begin{equation}
f({\rm u})=\left ( \frac{m_e}{2\pi k T_e}\right )^{\frac{3}{2}} \exp\left(\frac{-m_e{\rm u}^2}{2kT_e}\right)~~.
\end{equation}
for electron temperature $T_e$.
%and the thermal velocity is ${\rm u}_{th}=\sqrt{kT/m}$.
Similar analysis yields the absorption coefficient $\alpha_{\omega}$:
\begin{equation}
\alpha_{\omega} = -\frac{4\pi}{3c} \frac{\omega_p^2}{\omega^2} 
\int_0^{\infty} \nu_{en}({\rm u})\zeta(\nu,\omega)
\frac{\partial f({\rm u})}{\partial {\rm u}} {\rm u}^3 d{\rm u}
\end{equation}
where the plasma frequency is given by
$\omega_p^2 = N_e e^2/(m_e \epsilon_0)$
for electron number density $N_e$.
These coefficients are combined to form the source function
$S_{\omega} = (1/n^2) j_{\omega}/\alpha_{\omega}$
which is then integrated along a ray $s$ through plasma column to
the observer to determine the net intensity per unit 
radian frequency, or flux density $I_{\omega}$:
\begin{equation}
I_{\omega} = \int_0^{\tau_0} S_{\omega}(\tau)~e^{-\tau} d\tau
\label{Iomega}
\end{equation}
where the optical depth $\tau$ is defined 
by $d\tau = -\alpha_{\omega} ds$. Note that the absorption
coefficient is not necessarily positive definite: under 
conditions where $\frac{\partial f({\rm u})}{\partial {\rm u}} > 0$,
e.g., if there is an inflection in the electron velocity distribution
function, then $\alpha_{\omega}$ can become negative and stimulated
emission will cause the radiation to grow
with propagation distance.

The MBR flux density received by a ground detector
is estimated by integrating the intensity thus derived over the solid angle of the
receiving antenna beam to yield Watts per m$^2$ per Hz over a given frequency band. 
The minimum detectable change in flux density
for a radio antenna and receiver is~\cite{Dicke}
\begin{equation}
\label{deltaI}
\Delta I_{\omega,min}=\frac{k_BT_{sys}}{A_{eff}\sqrt{\Delta t \Delta\nu}},
\end{equation}
where $T_{sys}$ is the noise temperature of the receiver system, $A_{eff}$
is the effective area of the antenna, $\Delta t$ is 
the receiver sampling time constant, and $\Delta\nu$ is the receiver bandwidth.
After passage of the relativistic shower front,
the emission continues during the electron thermalization time, $t_{th}$.
This time is determined by both elastic and inelastic collisions of electrons
with air molecules \cite{thermal1,thermal2}, 
giving $t_{th} \simeq 10$~ns for dry air at 1 atmosphere. The received radiation 
continues during the time the shower remains in the field-of-view of the
antenna. At 4 GHz, for a $D$=1.8 m dish (giving a beamwidth $2\lambda/D \simeq 5^{\circ}$),
this time can be many microseconds for a distant shower.

Note that there is also a direct analog possible
between an optical fluorescence detector which used photomultiplier ``pixels'' to
image the two-dimensional projection of the optical fluorescence intensity
along the shower, and a radio dish which can use an array of focal plane
receiver feeds (each of which is effectively a pixel) to image the MBR
intensity along the same shower. Feed pixel-arrays are not so common in radio 
astronomy because of the success of radio interferometry, but they are
established technology where low-resolution bolometric imaging is important
(for example, in submillimeter radio astronomy).

Based on the parameters assumed above, we have numerically integrated the
flux density for air showers at a distance of 10~km, and we find that the
MBR floor level of emission, including the full suppression term  given in
equation~\ref{zeta1} above with $\xi=1$, gives an average detection energy threshold of
order $10^{19}$~eV. 
However, as we have already noted, estimates of the MBR emission of UHECR air showers using 
the standard thermal electron formalism here indicate
that the energy threshold for detectability of the emission may depend
strongly on the behavior of the modified suppression 
term $\zeta_c$, which is affected by several classes of 
non-equilibrium conditions that can obtain in an air shower plasma.
We address these in the next section.

\subsubsection{Departures from Equilibrium conditions.}
A summary of the conditions under which departures from the
MBR floor are expected is given in Table~\ref{MBRassump}. 
Each of these conditions may play a role in air shower emission,
and we discuss each of them in turn.

\begin{table}[htb!]
\caption{\it Assumptions for the standard MBR derivation 
compared to actual conditions in air showers.\label{MBRassump}}
\vspace{.1in}
\begin{center}
\begin{small}
\begin{tabular}{|p{1.5 true in}|p{1.5 true in}|}\hline
{\em Assumed Condition, standard MBR derivation} & {\em Actual conditions for air shower plasma}\\ \hline
Maxwellian (thermal) electron speeds &  Non-thermal, cascade power-law with high-energy tail \\ \hline
Isotropic velocity and momentum distribution & Linear ion-trails introduce first-order anisotropy\\  \hline
Time-stationary, in thermal equilibrium & Highly non-stationary, fast-transient relaxation \\  \hline
Collision frequency a weak function of electron speed & N$_2$ cross section a {\em strong} function of electron speed \\ \hline
\end{tabular}
\end{small}
\end{center}
\end{table}

\paragraph{Stimulated Bremsstrahlung.}
As an example of the departure from the assumptions regarding the
velocity dependence of the electron collision frequency, Figure~\ref{n2x}(top) 
shows the experimental electron-molecular nitrogen momentum transfer 
cross section $\sigma_M$ in the energy
range of interest. Most notable is the 2.3~eV resonance due to 
elastic collisions that lead to rotational excitation~\cite{N2x}.
This resonance region is in fact the complement of what is observed
in optical fluorescence--the energy transferred in this 
region of the cross section is released in part through optical
fluorescence, and the complexity of it is in part mirrored in the
structure of the optical nitrogen fluorescence transitions.
Such highly nonlinear changes in collisional parameters
with electron energy strongly depart from the
assumptions used above, and similar effects have been found 
to lead to stimulated emission even in highly collisional plasmas under
some conditions~\cite{DEC}. 

In fact, stimulated bremsstrahlung
emission from gas discharge plasmas was observed in the 1980's in
a series of experiments~\cite{StimBrem1,StimeBrem2,StimBrem3}. Necessary, though
not sufficient, criteria can be stated for stimulated bremsstrahlung emission:
\begin{equation}
-E \frac{\partial \sigma_M}{ \partial E} > \sigma_M
\end{equation}
for anisotropic electron distributions whose direction is parallel
to the direction of the electric vector of the propagating radiation,
and 
\begin{equation}
-E \frac{\partial \sigma_M}{ \partial E} > 2\sigma_M
\end{equation}
for isotropic electron distributions. Figure~\ref{n2x}(bottom)
evaluates this condition for the molecular nitrogen case
shown in Fig.~\ref{n2x}(top), and it is evident that both of the
conditions above are strongly satisfied in the neighborhood
of the resonance. Under such non-equilibrium conditions,
an electron population inversion in the ionized region is possible, and
this can lead potentially to stimulated emission~\cite{DEC}. 
Such inverted populations have been observed in discharge experiments
in molecular nitrogen plasmas~\cite{Inverted1,Inverted2}.

\paragraph{Oxygen attachment.}
Molecular oxygen has a momentum transfer
cross section significantly lower than that of nitrogen over this energy range,
as shown in Fig.~\ref{n2x}(top), and, after weighting for abundance, its
effect on the overall momentum transfer cross section in air is minimal.
Although O$_2$ does not contribute much to the thermalization
of electrons in air, it does however play an extremely important role
in removing free electrons once they have thermalized, since the
three-body attachment cross section to O$_2$ rises steeply at low
electron energies. In fact the attachment time scale for room
temperature electrons is comparable to the $\sim 10$~ns 
thermalization time scale for 
hot electrons in 1-atmosphere air~\cite{Raizer}. 
Once attached to ions, the electrons can no longer contribute
to the bremsstrahlung continuum radiation. And since
the initial spatial distribution of the oxygen ions is 
highly structured, and the ions are almost stationary in the
short period during electron attachment, they impose 
a rapidly developing anisotropy in the {\em removal} of
electrons from the free distribution, creating MBR in the
free-bound transition. This effect also imposes anisotropy in
the velocity distribution of the electrons.

\paragraph{Plasma correlations.}
The Debye length $\lambda_D$, over which an electron is fully shielded from
a neighboring ion in a plasma, is given by 
\beq
\lambda_D = \sqrt{\epsilon_0 k T_e/(e^2 n_e)}
\label{debye}
\eeq
where $n_e$ is the electron number density.
For an EAS plasma
at an energy of 10 EeV or more, $\lambda_D \sim 1-2$~cm. Since
over this distance there may be several hundred ion pairs along each
relativistic through-going particle track, along with several hundred 
to several thousand tracks per square cm in the vicinity of the
EAS core, electrons do not behave entirely independently but
are subject to weak bulk plasma effects at some level which will 
produce phase-space correlations. For our case we may class
these effects together with the attachment effects above; in either
case the end result will be parameterized via equation~\ref{zetac}.

\subsubsection{Radiative Coherence.}

\label{coherence}

While the field strength for a single electron is accurately described
by the MBR theory, the summation of these fields in the presence of
correlated velocities can significantly alter the resulting ensemble
field strength. Such alterations, which may be produced by
intrinsic shower geometry, or electron-photon correlations from
stimulated emission, or by other plasma effects, 
still may be described via simple vector sums
of the field strength of each of the radiating particles involved.

For individual emitters the resultant field strength will
grow as a phasor sum~\cite{Goodman}:
\beq
\vec{E}~=~ \sum_{j=1}^{N_e} \vec{\epsilon_1}({\vvec}) \exp{(-i~\vec{k}\cdot\vec{x}_j)}
\eeq
where $N_e$ is the total number of electrons in the plasma, 
$\vec{\epsilon_1}({\vvec})$ is the field radiated from a single electron,
$\vec{k}$ is the wave vector of the radiation, and $\vec{x}_i$ is
is the position of the $j$th electron with respect to the observation
point.  The total radiated far-field power per unit area 
$P/A$ is given by  the magnitude of the Poynting flux
\beq
P/A = |S_{tot}|~=~ |\vec{E}|^2 / Z_0
\eeq
where $Z_0\simeq 377~\Omega$ is the impedance of free space. In the limit
of complete coherence, the phase factors $\vec{k}\cdot\vec{x}_i$ are all unity,
$|\vec{E}|~=~N_e~\epsilon_1$, and the total coherent power is $P_{coh}~=~N_e^2~P_1$, where
$P_1$ is the power radiated from a single electron. Since
$N_e$ is proportional to shower energy, the coherent power depends 
quadratically on the energy of the primary particle.
In the incoherent limit, the sum of the phase factors corresponds to
a two-dimensional random walk in the real and imaginary components
of the resultant field strength, and the total power grows as
$P_{incoh}~=~N_e~P_1$. 

While in general the partially coherent case requires a
detailed knowledge of the electron phase space distribution
function,  we can get a qualitative sense of the behavior by considering a
case where the $N_e$ electrons consist of $M$ subgroups of $\mu_e$
electrons each, such that $N_e=M\mu_e$. Assume that the $\mu_e$
electrons in each subgroup radiate coherently, but that 
the subgroups themselves are uncorrelated. Thus, while
the radiated fields from the $M$ subgroups add incoherently,
the subgroup electrons themselves radiate coherently, and
the resulting {\em partially} coherent power is
$P_{part}~=~ M~\mu_e^2~P_1$, now quadratic in $\mu_e$ rather than
$N_e$. The ratio of the partially coherent power to the incoherent
power is proportionally
\beq
\frac{P_{part}}{P_{incoh}}~=~ \frac{M~\mu_e^2~p_i}{N_e~p_i} = ~\mu_e.
\label{partcoh}
\eeq
Similarly the ratio of coherent-to-incoherent power grows as $N_e$. 
Since the plasma density of ionization electrons in a shower
scales linearly with shower energy, both the coherent and partially
coherent regimes will yield radiated power that grows {\em quadratically} with
shower energy. In fact, as soon as $\mu_e \geq 10$, coherence begins to dominate
over the incoherent component by an order of magnitude or more. 
Even modest correlations among the
shower ionization electrons can thus rapidly lead to much larger detected emission
than expected. 

We have parameterized these effects using the correction term $\xi$ which
modifies the collisional decoherence factor $\zeta$ as described
above. In practice empirical data will be required to establish
the emission constants associated with these factors, as is
the case for all other emission mechanisms in a real air shower.

\subsection{Accelerator beam tests.}

Motivated by the fact that even the floor level of 
fully-suppressed emission
from the MBR process appeared to us to be detectable under
air shower plasma conditions,
we have performed two accelerator tests 
designed to measure the MBR in a laboratory air-shower
plasma. In these experiments we have found good evidence 
for microwave continuum emission with
characteristics suggestive of a major departure from the standard
incoherent MBR emission scenario, not an unexpected result given
the variety of different non-equilibrium, non-thermal, and partially
coherent processes that are possible.
We detail these results here.
\begin{figure}[hbt!]
  \centerline{\includegraphics[width=3.2in]{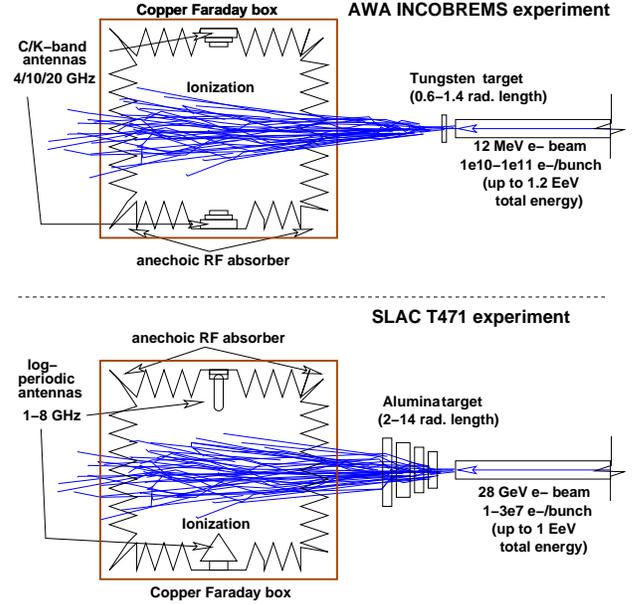}}
  \begin{small}
  \caption{ \it Schematic of AWA INCOBREMS (top) 
  and SLAC T471 (bottom)
  experiments, which used electron beams to shower
  in either Tungsten or alumina targets to produce 
  ionization inside an anechoic Faraday chamber,
  observed by internal antennas.
  \label{bremsetup}
  }
  \end{small}
\end{figure}
\begin{figure}[hbt!]
  \centerline{\includegraphics[width=3.2in]{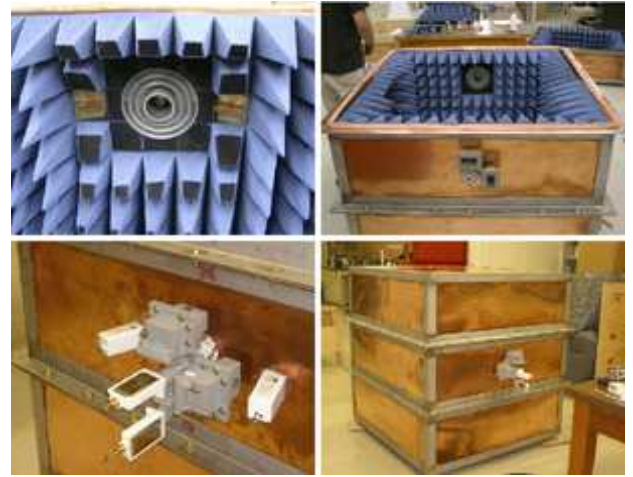}}
  \begin{small}
  \caption{ \it Views of the exterior and interior of the Faraday
  anechoic chamber used for measurements of microwave continuum
  emission in the INCOBREMS and T471 experiments. The box
  is approximately a 1~m cube in dimensions.
  \label{anlbox}
  }
  \end{small}
\end{figure}
\begin{figure}[hbt!]
  \centerline{\includegraphics[width=3.3in]{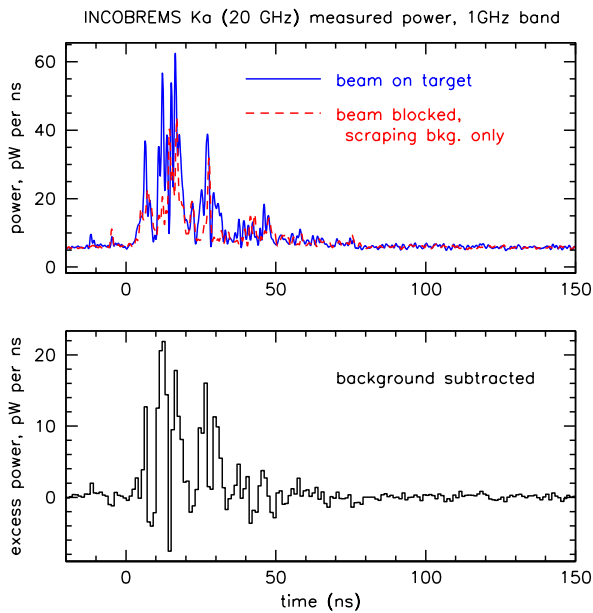}}
  \begin{small}
  \caption{ \it
Phase-stable (partially coherent) component for 20~GHz emission observed from the
5mm tungsten target, where background subtraction of beam-scraping backgrounds was
possible. Upper pane shows 
blue as total emission and red as the background due to stray ionization from the beam;
the lower pane show the background-subtracted results. The data is averaged over
several thousand beam shots.
  \label{fig:awa1}
  }
  \end{small}
\end{figure}

\subsubsection{AWA INCOBREMS.}

In June 2003, the INCOBREMS experiment was performed at the Argonne Wakefield 
Accelerator (AWA).  The beam used for the experiment consisted of 12~MeV electrons, pulsed
in $\sim$7~nC charge bunches of 1.2~cm longitudinal thickness 
(40~ps duration) containing $\sim\!\sci{4}{10}$
electrons. This gives a typical total beam energy of $5 \times 10^{17}$~eV. 
The beam was collided into a fixed radiator of tungsten
with a thickness of 2 or 5~mm 
(tungsten has a radiation length of 3.5~mm), producing a core of photons with
energies 5--10~MeV as well as some lower energy electrons. 
Typically 40-90\% of the total energy
was extracted into photons that traversed the chamber, depending on the radiator.
The conversion was necessary since the 12~MeV electron energy was well
below the critical energy in air, and thus inadequate to
produce a fully developed shower. By converting to gamma-rays we avoided a large
excess negative charge associated with the passage of the electron beam through the
Faraday chamber. The photons
entered an air-filled $\sim$1~m$^3$ copper anechoic Faraday chamber
which prevented interference from outside electromagnetic radiation
and absorbed transition radiation caused by beam effects on the
copper. C (3.4--4.2~GHz), Ku (10.7--11.8~GHz) and Ka (20.2--21.2~GHz)
band commercial radio receivers were mounted on the insides of the
chamber to measure subsequent radio emissions. 
Figure~\ref{bremsetup}
shows a schematic view of the general layout for both this and the 
subsequent SLAC experiments,
and  Fig.~\ref{anlbox} shows several views of the anechoic Faraday chamber
employed in both cases (here shown with the antennas and receivers used
for the INCOBREMS experiment).

The photon bunches in the AWA experiment typically deposited about  
1~PeV of ionization energy while passing through the Faraday chamber.
Since the radiation length of electromagnetic particles in air is of order
300~m at sea level, the deposited energy is of order 
$$E_{chamber} \sim \frac{1}{300}(1-\frac{1}{e})(5 \times 10^{17}~{\rm eV}) \simeq 10^{15}~{\rm eV}.$$
The mean energy required per ion pair is of order 30~eV, and there are thus
about $3 \times 10^{13}$ ion pairs produced in the chamber for each beam shot.
The distribution of the free electrons in the plasma is of course much denser than
the equivalent PeV cosmic ray air shower. 
Most ($\sim 80$\%) of the plasma is produced in a central
cylindrical region through the chamber, with a radius of order 25~cm, and a mean
plasma density of order $10^8$~e$^-$ cm$^{-3}$.

Based on the expectations
of MBR, we expected to observe emission that was incoherent, with intensity that
scaled linearly with beam current. When our initial observations indicated that 
the emission appeared to be scaling coherently, with intensity going as the square
of beam current, we developed analysis methods that 
attempted to separate the two components, taking advantage of the fact that the
phase stability of the coherent component allows for it to be subtracted from the total
emission. (We initially adopted
the term ``phase stable'' to describe this emission, since the degree of 
coherence was unknown.) To attempt to separate out the various components, we
used two thicknesses (2mm and 5mm) of tungsten radiators to convert the electron
beam to bremsstrahlung photons, since this provided a different bremsstrahlung
energy distribution and beam emissivity, which could modify the relative
contributions of the two components. Our measurements were unable to separate
out any significant incoherent signal component in these data, due primarily 
to background issues, and indicating that the coherent or phase-stable
component was at least an order-of-magnitude larger.

In Fig.~\ref{fig:awa1} (top), 
we show results of the AWA measurements at 20GHz (Ka-band) using the
5~mm radiator, which gave the cleanest background-subtracted results.
Partially coherent emission was observed about 50~ns after beam passage. 
While the presence
of the beam gave clear excess power levels, there is also considerable
apparent ``signal'' present when the beam was blocked with lead just before
our system. We found that acceleration and production of the
electron beam within the of order 15~m length of this accelerator required that
a major fraction (80\% or more in some cases) of the electrons were removed
by upstream collimation (a controlled scraping of the beam), 
but without any way to remove the secondary
radiation (mostly hard bremsstrahlung photons) that this produced.
This led to a high level of background ionization in both our chamber
and the surrounding vault, leading to doubts
about the reliability of the results.  This is evident in Fig.~\ref{fig:awa1},
where the backgrounds with the beam blocked can be seen to at times exceed
even the apparent signal. We confirmed the presence of such backgrounds
using external ionization detectors. We also checked carefully 
whether any portion of these
backgrounds could be due to radio-frequency interference, and we confirmed 
that this was not the case. 

However, we note that our conclusion regarding these
backgrounds implies that they are actually stray signal, due to the unwanted beam
albedo (the beam components that caused scattered bremsstrahlung due to impacts
with the side-walls of the beampipe). Thus it appeared to us that the presence
of microwave emission from the chamber ionization was unavoidable. To
further pursue the investigation with a more tightly controlled beam,
another experiment was scheduled at The Stanford Linear Accelerator Center.

\begin{figure}[hbt!]
  \centerline{\includegraphics[width=3.2in]{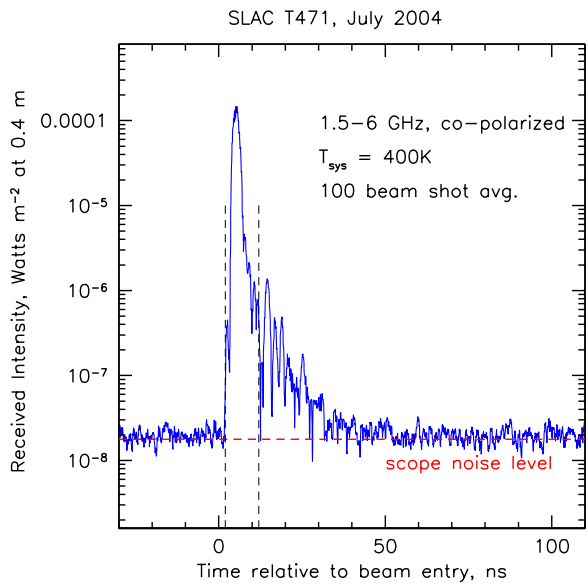}}
  \begin{small}
  \caption{\it 
  Average microwave emission amplitude from 100 beam shots taken
  near shower-maximum in the 2004 SLAC T471 experiment, using
  a broadband antenna that was polarized along the electron beam axis, 
  and was thus sensitive to partially coherent radiation directly from the
  relativistic electron shower as it transited the Faraday
  chamber. A strong initial pulse is seen, with rapid decay,
  followed by a second exponential tail with a longer decay. The noise level
  is in this case determined by the limited dynamic range of
  the oscilloscope used, rather than the thermal noise level.
  \label{LPDA_copol}
  }
  \end{small}
\end{figure}
\begin{figure}[hbt!]
  \centerline{\includegraphics[width=3.2in]{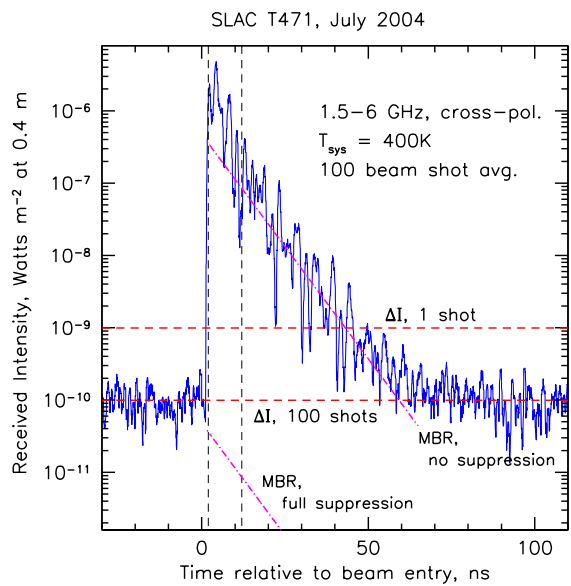}}
  \begin{small}
  \caption{\it 
  A plot similar to the previous figure, but now using a cross-polarized
  antenna which was insensitive to radiation polarized with 
  the electron beam. The dynamic range of the system was now improved
  so that the noise level is determined by thermal noise, and the detected
  microwave emission extends out to 60 ns or more, with an exponential
  decay time constant of about 7~ns. The upper and lower dashed red
  horizontal lines indicate the minimum detectable intensity, as given
  by equation~\ref{deltaI}, for the single-shot case, and the
  100-shot average. The diagonal dot-dash lines are the two extreme-case
  estimates for MBR emission: the upper case for no net
  collisional suppression and the lower case for maximal collisional
  suppression of the emission, both for the case where the electron thermalization
  time constant is the source of the 7~ns exponential decay observed.
  \label{fig:slac04}
  }
  \end{small}
\end{figure}

\subsubsection{SLAC T471/E165.}

In the following year, a similar experiment, T471, was performed at the
Stanford Linear Accelerator Center.  The configuration of this experiment was
largely the same as that of INCOBREMS, but additional precautions were taken
against EMI and beam backgrounds, and verified in lab and beam calibration tests. 
This experiment was coordinated to be operated just downstream of the
E165 FLASH experiment, which was used to do precise calibration of 
air fluorescence for the HiRes collaboration~\cite{FLASH}.
The SLAC T471/E165 experiments also used a precisely controlled,
28~GeV electron beam which was collided with a target consisting of 90\% Al$_2$O$_3$
and 10\% SiO$_2$ to make showers with varying particle number, from 0 to 14
radiation lengths of material. In T471/E165, the 28.5~GeV 
electron bunches were used directly
to create the showers with no intermediate conversion to photons via
a bremsstrahlung radiator, as this was unnecessary given the high electron energy.
Bunches with a typical charge of $\sim 2 \times 10^7$ electrons were used, giving a total shower
energy of typically $6 \times 10^{17}$~eV, very similar to
those used at AWA.

\begin{figure}[hbt!]
  \centerline{\includegraphics[width=3.25in]{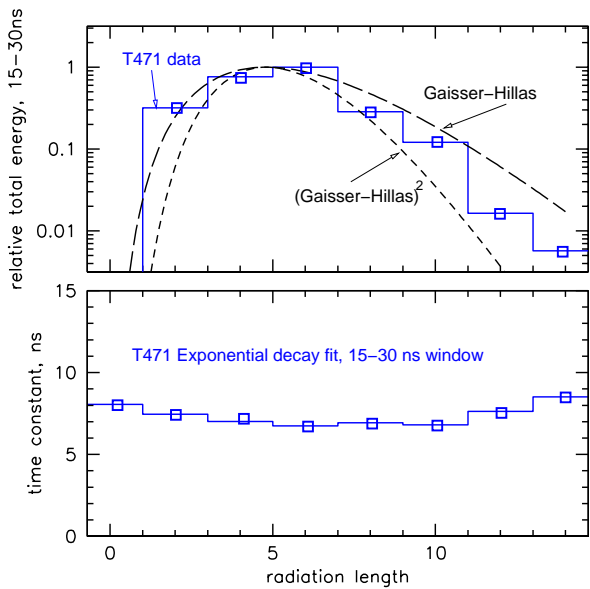}}
  \begin{small}
  \caption{ \it
Top: relative integrated energy in the tail of the microwave emission,
between 15 to 30 ns, as
a function of the depth of the shower in radiation lengths. The curve shows
a Gaisser-Hillas shower profile for comparison, which peaks at about
4.7 radiation lengths at shower maximum. By contrast, the microwave tail emission
shows some early radiation probably due to the initial electron bunch.
Near shower maximum, the shower charge overcomes the beam charge, and the
emission appears to follow the shower profile near shower maximum, though it falls below the
particle number profile at late times. 
  \label{lpdafits}
  }
  \end{small}
\end{figure}

Figure~\ref{LPDA_copol} shows 
results from measurements of the emission over the
1.5-6~GHz band, using an antenna that was co-polarized with the 
electron shower momentum. Here the square of the average signal voltage is plotted
vs. the time after beam entry into the Faraday chamber. The transit time
for the chamber is about 3.3~ns for the beam. An initial strong impulse
is observed at the first causal point in time after beam entry. This impulse
is found to be highly polarized with the plane of polarization aligned
with the beam axis and Poynting vector, characteristic of transition and
radio Cherenkov radiation. Such emission was anticipated, and is damped almost
immediately due to the microwave absorber ($\geq 30$~dB per reflection even
at angles of order $55^{\circ}$ from normal incidence) that covers the interior of the
Faraday chamber (the implied average time constant for 
quasi-exponential decay of reflections is
of order 1.3~ns for this absorber in this geometry).  The noise
level in this plot is dominated by digitization noise, since the sensitivity
had to be reduced in order to achieve enough dynamic range to see the
strong initial impulse.

In Fig.~\ref{fig:slac04} we plot the same measurements made with a 
cross-polarized antenna, which was therefore insensitive to the 
relativistic shower emission, with a 20~dB cross-polarization rejection
factor. In this case the strong initial impulse
is not prevalent though in fact the leading edge is likely to be slightly influenced by
the -20dB leakage from the other polarization. 
The exponentially-decaying tail of emission extends out
to 60~ns or more, with noise levels now determined by the thermal noise level
rather than scope noise. Based on 
beam-out vs. beam-in and beam-on vs. beam-blocked measurements, no 
beam-related background (either ionization or EMI) was present, and thus no background
subtraction was necessary. Several curves are also plotted with the data.
The horizontal lines indicate the thermal noise level for single shots and
for the average of the 100 beam shots used here, based on 
equation~\ref{deltaI} above. The diagonal 
dot-dash curves are model predictions based on equation~\ref{Iomega} above,
calculated for the two extreme cases of the collisional term $\zeta_c$
from equation~\ref{zetac}, one for the case of no suppression ($\zeta_c=1$)
and the other for full collisional suppression ($\xi=1$). It is evident that, if
MBR is responsible for this emission, the collisional suppression is
almost completely offset by the partial coherence.

In Fig.~\ref{lpdafits}(top) we plot the behavior of the integrated microwave
energy in the 15-30 ns window as a function of shower depth in radiation lengths.
The emission from the direct beam, which adds noticeably to the shower emission up to
about 4 radiation lengths, has been subtracted here in proportion to the
depth in radiation lengths, so that the contribution from the shower 
emission alone can be compared to expectations. The upper (long-dashed) 
curve shown is a standard
Gaisser-Hillas profile~\cite{Gaisser77}, peaking at about 4.7 radiation
lengths for these showers. The lower (short-dash) curve is a Gaisser-Hillas
profile, but now scaled as the square of the particle number in the shower.
It is evident that the shower emission scales
roughly with the particle number in the shower, but appears to drop below the
standard Gaisser-Hillas profile at large shower depths, although not enough
to warrant scaling that is quadratic in particle number. This behavior provides evidence
that the process for the emission is relatively insensitive to the plasma
density. At larger shower depths in particular the plasma density decreases by
1-2 orders of magnitude with only factors of 2-3 apparent drop in the relative
microwave emission compared to expectation based on the standard Gaisser-Hillas
profile.

Fig.~\ref{lpdafits}(bottom)
shows that the fitted time constant of the decay of the emission power is roughly
constant with shower depth at about 7~ns, with some indication that it may be increasing for
large shower depths. The near constancy of this parameter indicates that
the underlying physical process that removes the radiating electrons from
the emitting populations is nearly completely insensitive to plasma
density.

\begin{figure}[bht!]
  \centerline{\includegraphics[width=3.2in]{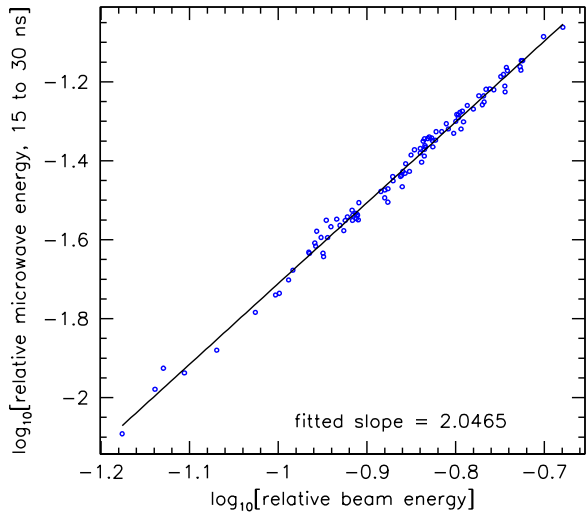}}
  \begin{small}
  \caption{\it
   Plot of relative microwave energy in the tail of the observed
  air plasma emission vs. the relative microwave energy observed 
  in an external transition-radiation
  monitor of the beam current, which is directly proportional to beam
  energy. The observed microwave power follows closely
  a quadratic rise with beam energy, characteristic of coherent radiation.
  \label{t471_coh}
  }
  \end{small}
\end{figure}

The radiation observed in T471 is also partially coherent.
This is shown in Fig.~\ref{t471_coh}, which plots the 
integrated microwave power from 15-30 ns after the main pulse
vs. beam energy as measured by an external transition radiation detector.
The quadratic correlation here indicates 
that the partially coherent portion of the emission
dominates completely over incoherent emission. The coherent 
emission appears to be several thousand times the 
expected incoherent emission level, implying that subgroups of
$\mu_e \simeq 10^{3-3.5}$ electrons are radiating quasi-coherently within their
subgroup, using the notation of equation~\ref{partcoh} above.
Given that the showers used in T471 created $3 \times 10^{13}$
ionization electrons within the Faraday chamber used, 
the net correlation of $\sim 10^{-10}$
is still extremely small, and it is evident that
this level of partial coherence is very far removed from the 
full-spatial coherence that obtains in coherent Cherenkov 
or transition radiation. 

We note that the Debye length (equation~\ref{debye} above) for the T471 plasma
is initially of order 2~mm when $T_e \simeq 10^{4.5}$~K, and within one Debye radius there are
of order $10^7$ free electrons initially. Thus a weak correlation 
of $\sim 0.01$\% within a Debye radius appears to be all that is required to create
the observed partial coherence effects. This analysis does not account for
the rapid evolution of the Debye length as the electrons cool however. At $T_e=10^3$~K,
close to ambient, $\lambda_D \simeq 0.3$~mm, and the Debye volume then contains
of order $10^5$ electrons, still requiring only a 1\% correlation coefficient. 
However, any prediction using a simple plasma-correlation model requires understanding
of the dynamics of the cooling event before a self-consistent picture can emerge.

\subsubsection{Scaling to air showers.}

Under the assumption that standard
radiation scaling laws obtain,  we can make an estimate of
the threshold for detectability of the emission seen in Fig.~\ref{fig:slac04}, 
if we scale it to air shower observation distances and a realistic detection
system.

\begin{figure*}[ht!]
\includegraphics[width=5.3in]{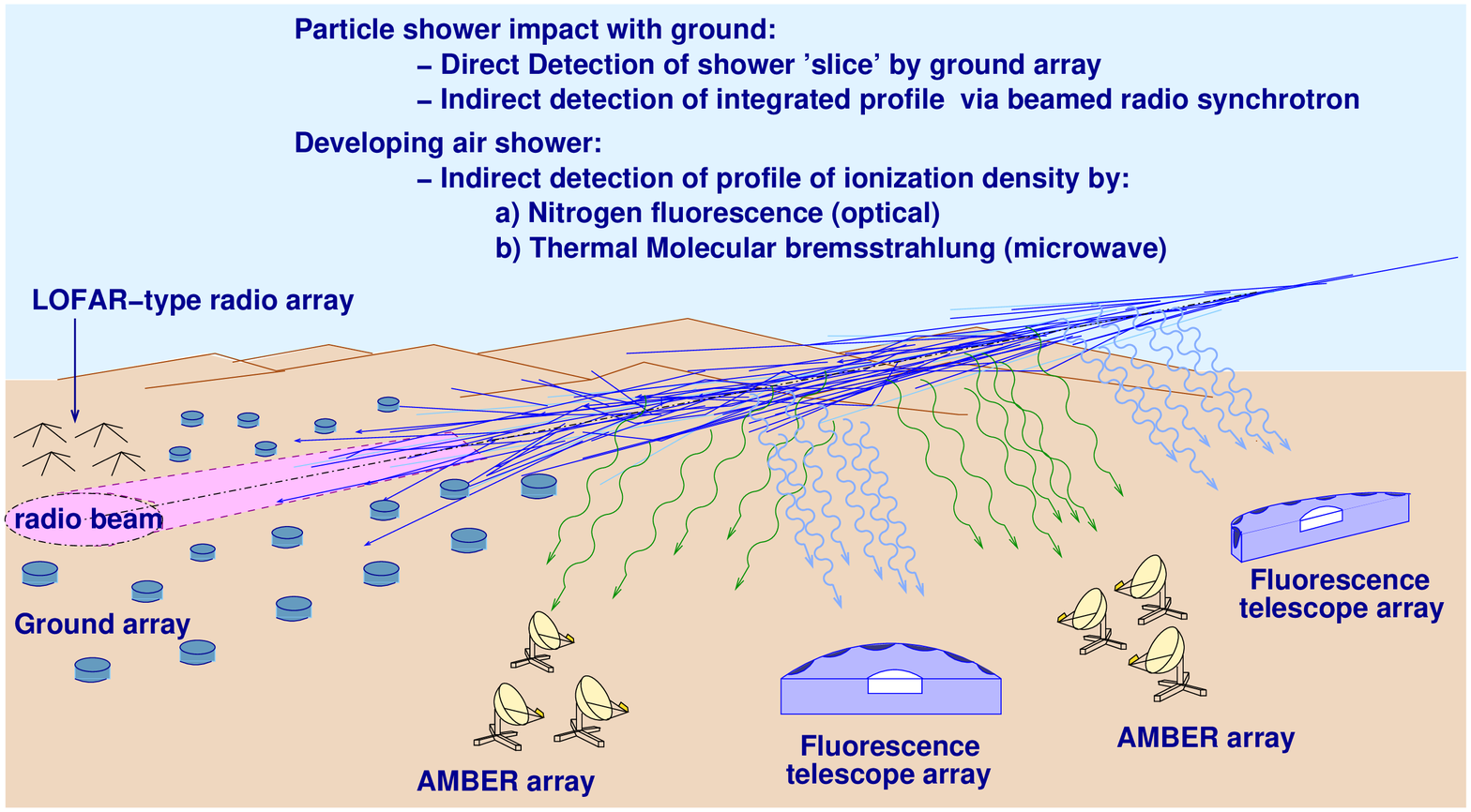}
\caption{\it Conceptual sketch of how  microwave bremsstrahlung
detection relates to other methods of ultra-high energy 
cosmic ray air shower detection. \label{Ambergeom}}
\end{figure*}

\begin{table*}[htbp!]
\caption{
Detectability of air showers assuming that the T471 
shower emission shown in Fig.~\ref{fig:slac04} can be scaled to an
air shower at 10~km distance, reaching maximum at 5~km altitude,
and detected with a 1.8~m diameter dish at 4~GHz (C-band) using standard
satellite dish technology. Results are estimated both for linear
and quadratic scaling with shower energy.
  \label{MBRtable}
  }
 \vspace{1.5mm}
  \centerline{\includegraphics[width=6.25in]{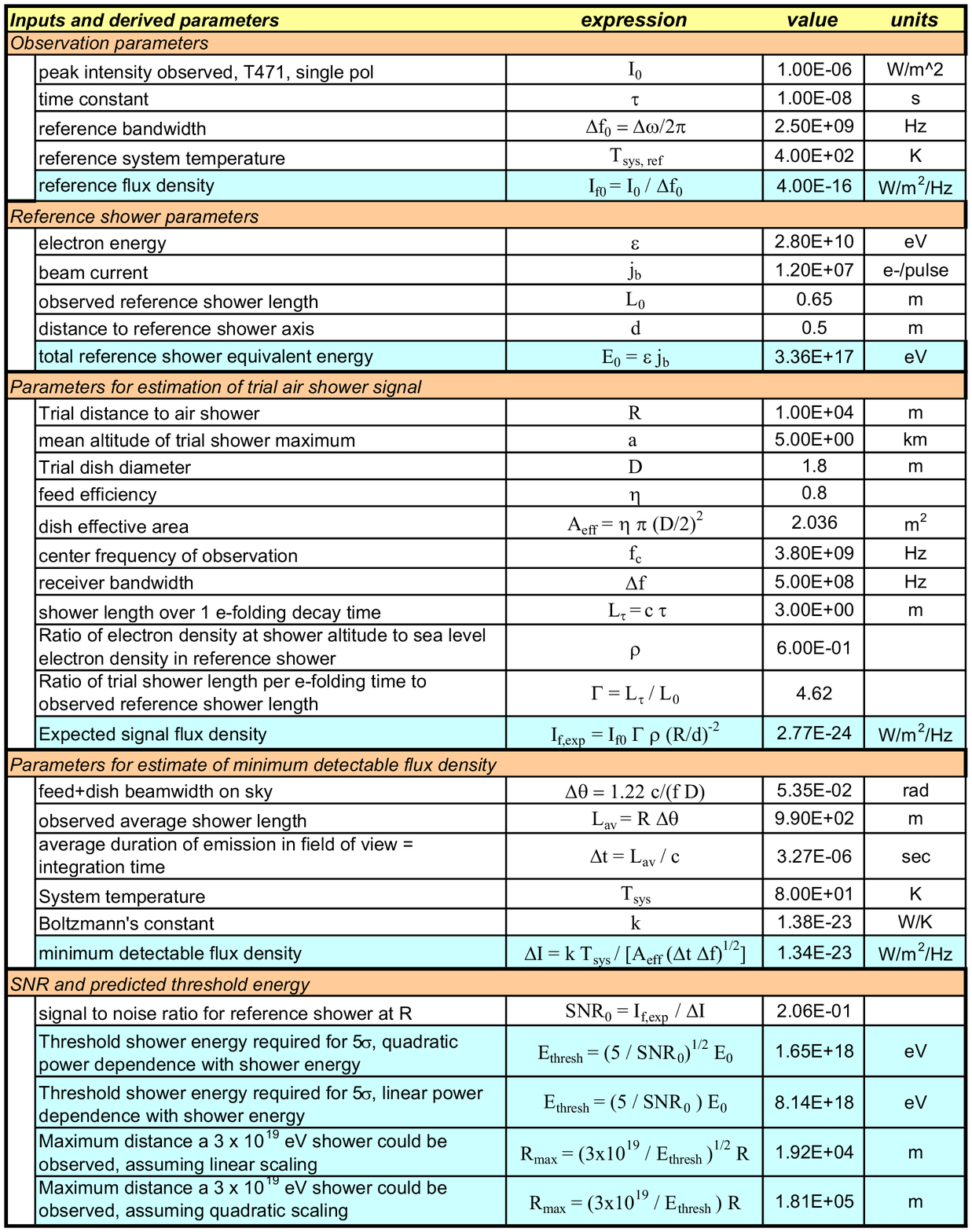}}
\end{table*}

\begin{figure}[htb!]
 \centerline{ \includegraphics[width=0.9\columnwidth]{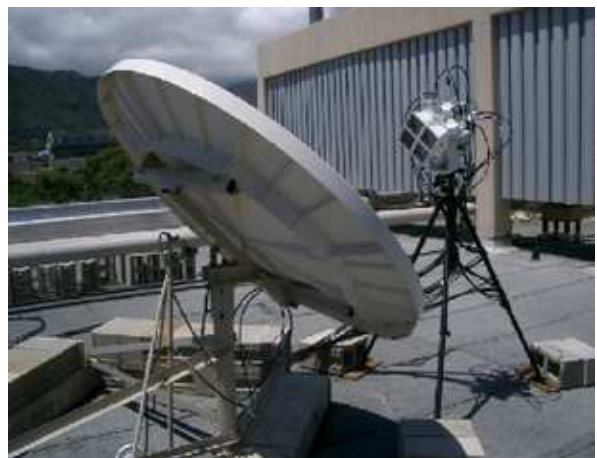}}
  \centerline{\includegraphics[width=0.9\columnwidth]{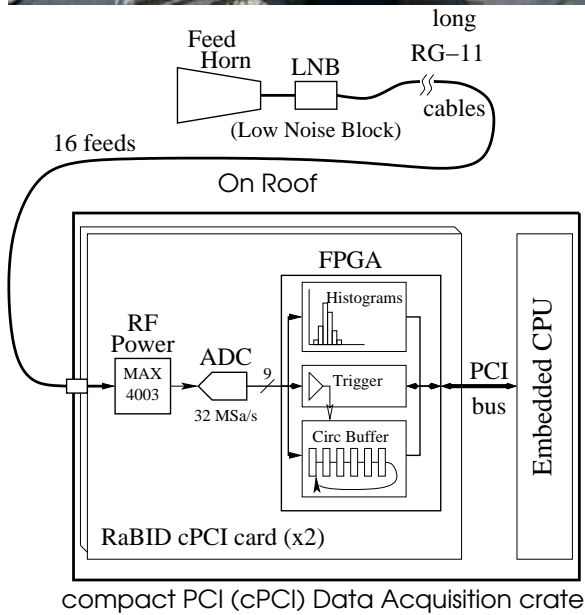}}
  \caption{\it Top: Prototype AMBER telescope and feed array on the roof of the
physics building  at UHM.\label{fig:rabidpic1}
Bottom: AMBER detector readout chain.  The feed horn
signals are amplified and down-converted in a Low Noise Block and
then transmitted to a pair of RaBID cards for processing.  See text
for details. \label{fig:rabid_scheme}}
\end{figure}
\begin{figure}[htb!]
  \includegraphics[width=0.97\columnwidth]{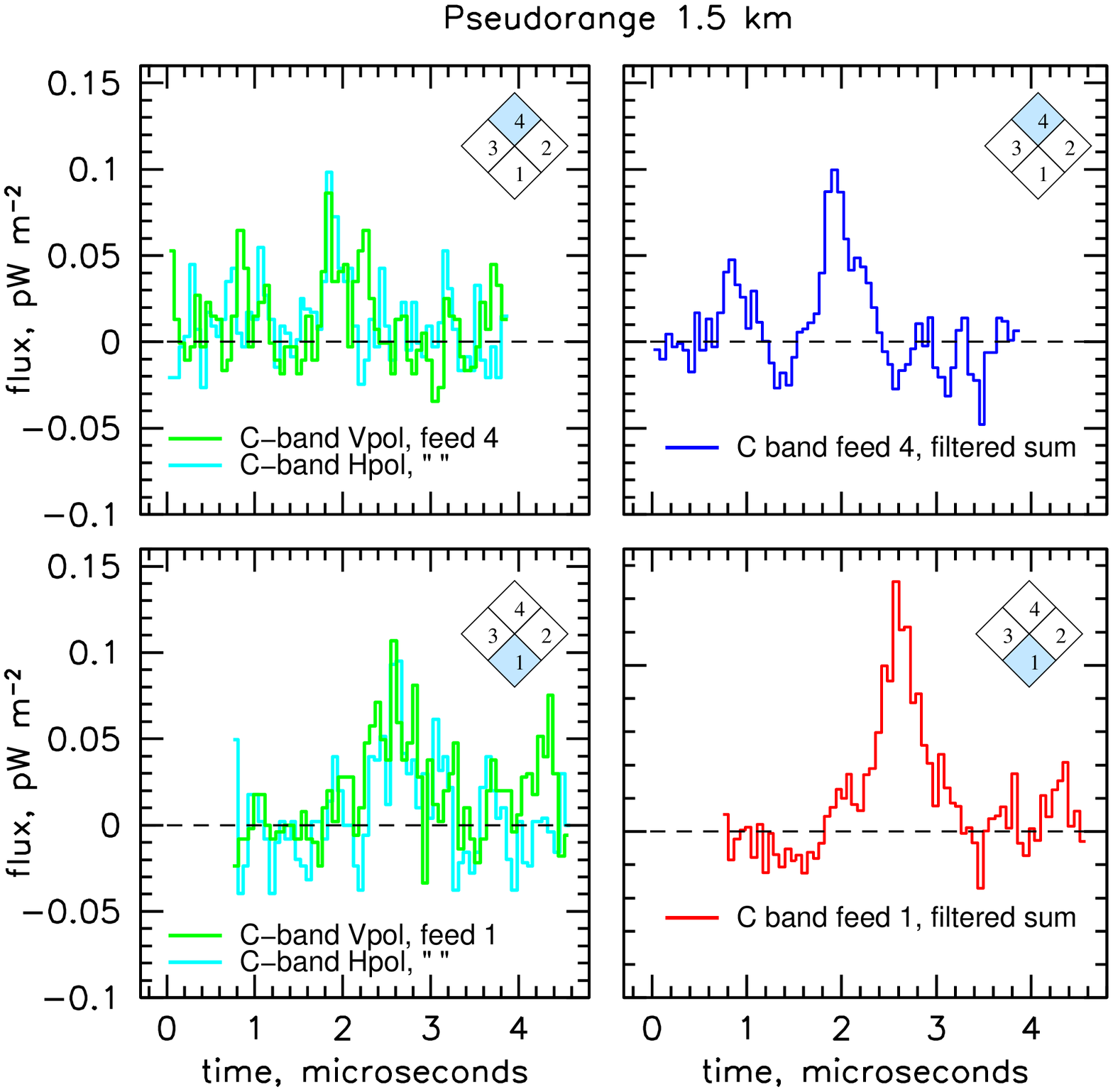}
  \begin{small}
  \caption{\it
Example of an event recorded recently with
the prototype RaBID system in Hawaii. The two events occurred
sequentially in the two feeds noted, and triggered both
polarizations at both feed 4 (top) and feed 1 (bottom), creating a downgoing
event trigger. Left,: Raw data, showing peaks in both H and V
polarizations for the C-band feeds. Right: Signal after co-adding
the polarizations and applying Wiener filtering to remove the
high frequency thermal noise fluctuations. The pseudorange is
based on the 680 ns dual-feed crossing time. 
  \label{Amber_ev1}
  }
  \end{small}

\end{figure}
To do this, we take the flux density as estimated from the data in Fig.~\ref{fig:slac04},
using the weighted average T471 antenna effective area of 0.05~m$^2$. Using
this flux density and the equivalent shower energy of $3.4 \times 10^{17}$~eV,
we scale to an equivalent air shower at a distance of 10~km.
%Table~\ref{MBRtable} gives the parameters and dependencies of one such
%realization of this scaling, based on the AMBER unit dish system we
%will detail in the following section. 
The scaling corrects for the lower
electron density expected for a typical 5~km air shower altitude. We also
assume an integration time (several hundred ns) 
based on angular transit times for showers at roughly this distance,
with the peak flux density determined by the emission over an
interval comparable to the thermalization time. We consider both 
linear and quadratic scaling of
the emission with electron density, and with regard to the shower geometry,
as long as the transverse diameter of plasma column is contained within the antenna
beam, we assume there is a direct scaling from the T471 observations to air shower
observations.

The results of this analysis indicate that, if the partially coherent emission
observed in T471 scales only linearly with shower energy (as might be expected
in the pure ``incoherent'' case), then the threshold of the AMBER system at 10~km
is of order $8 \times 10^{18}$~eV in shower energy. If the scaling is quadratic with shower
energy, as the data suggests, 
the threshold is lower, of order $1.6 \times 10^{18}$~eV. 

We can also estimate the maximum distance to which a shower at the
GZK threshold energy of $3 \times 10^{19}$~eV could be observed under these
same conditions: for the linear-scaling case, the distance is of order 20~km;
for the quadratic case it is much larger, of order 200~km, but of course
in this case earth-curvature and atmospheric attenuation would also require
consideration. In either case, the current emission parameters strongly warrant
further investigation of the potential for development of MBR detection
of air showers.

As noted in a previous section, an air shower plasma can also be characterized in
terms of its Debye length. For a $10^{19}$~eV shower, with an initial electron 
density of order $10^{10}$ e$^-$ m$^{-3}$ within a few m of its core, 
$\lambda_D \simeq 7$~cm for $T_e = 10^4$~K during the early period of the electron 
cooling, and $\lambda_D \simeq 1$~cm once the electrons have cooled close to ambient levels.
The electron number within a Debye radius similarly evolves from $\sim 10^7$ to $\sim 10^4$
over the same cooling period, several tens of ns. These values leave open the possibility
of correlated electron behavior comparable to those seen in our T471
experiment, if the correlations are related to plasma density parameters.

\subsubsection{Beyond detection: Shower Calorimetry with an AMBER array.}

%\subsubsection{Potential Impact of Microwave EAS Detection.}

The importance of MBR detection of EAS rests in the potential
that it will yield the observational advantages comparable to those
of optical fluorescence without the shortcomings associated with weather
and limited duty cycle.  By observing 
MBR, one is observing an EAS from the same perspective
as with optical fluorescence, via energy-loss processes that are intimately related
to the excitation of molecular nitrogen that leads to air fluorescence.  
However, observations can occur 24 hours per day,
and at the microwave bands of interest there is virtually no attenuation due to
atmospheric contamination from aerosols or clouds. Even heavy rain
leads to attenuation of$\leq 1$~dB above elevation angles of
30\deg at C-band (4-6~GHz), a 20\% effect. Initially, while the MBR technique 
is being cross-calibrated with respect to an optical fluorescence and
ground array, this immunity to atmospheric effects can yield 
immediate benefits in helping to extrapolate the energy scale for
distant events, where optical fluorescence is most affected by
aerosols and other atmospheric uncertainties.

Commercially designed microwave 
reception equipment can be easily weatherproofed,
and future arrays
would most likely be able to employ off-the-shelf satellite television components,
taking advantage of the tremendous economy of scale in wireless and
satellite television technology. Following validation of the technique
in coincidence with an existing EAS installation, MBR detectors could
be potentially deployed as standalone UHECR observatories.
Critical to the success of such an observatory is the ability of
MBR to do precision shower calorimetry. There appears to be good
initial evidence from the T471 experiment that such calorimetry
can be done with precision that is comparable to current techniques.

Referring to Figure~\ref{t471_coh}, we stress that individual single-shot
measurements of the integrated microwave energy in the shower
can be used to determine the beam energy in the experiment 
to about 2-3\% precision, once the overall energy scale is set (in this
case by an external beam current monitor). This level of precision is
set entirely by the instantaneous signal-to-noise of the data,
and is not systematics-limited, as evidenced by the precise quadratic
scaling observed. In the T471 experiment, this scaling was observed
over more than a decade (not shown in Fig.~\ref{t471_coh}) of 
energy. In addition, the same scaling was seen at many different
relative plasma densities (created by sampling the shower at
different depths of development) in the experiment, indicating that it
is not dependent on shower age. Such results give us good
confidence that, with sufficient attention to careful detector and
system design, and adequate calibration, an MBR observatory could 
provide shower calorimetry which was comparable to that
of existing techniques.

\begin{figure*}[htb!]
  \includegraphics[width=4in]{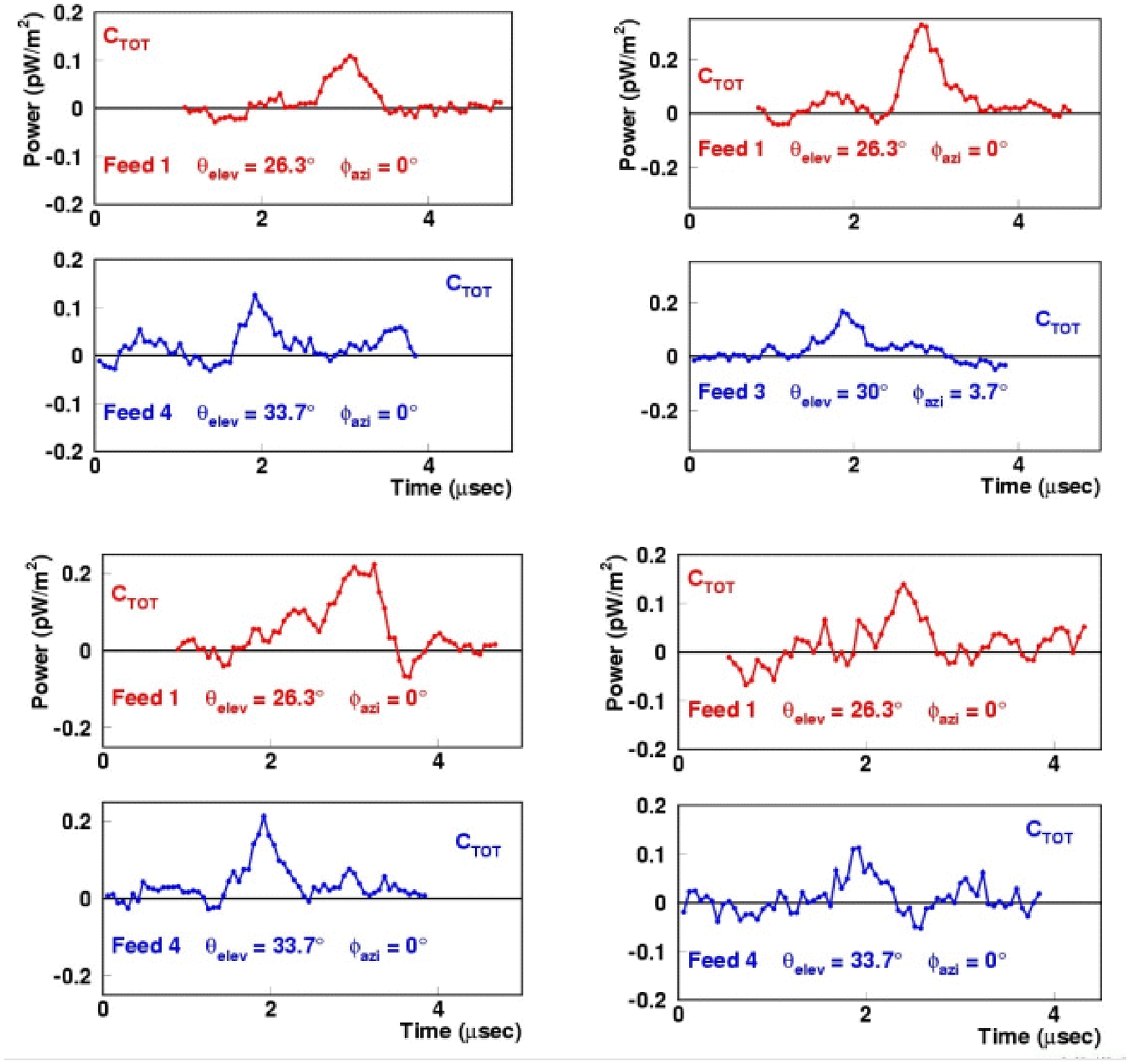}
  \begin{small}
  \caption{\it
Examples of other events recorded by the prototype AMBER system,
meeting the criteria for EAS candidates. (In these data, the
convention is reversed compared to Fig.~\ref{Amber_ev1},
with the lower curve for each event originating from a feed
which scans a higher elevation in the sky.)
  \label{amber_evs}
  }
  \end{small}
\end{figure*}

\begin{figure}[htb!]
  \includegraphics[width=3.in]{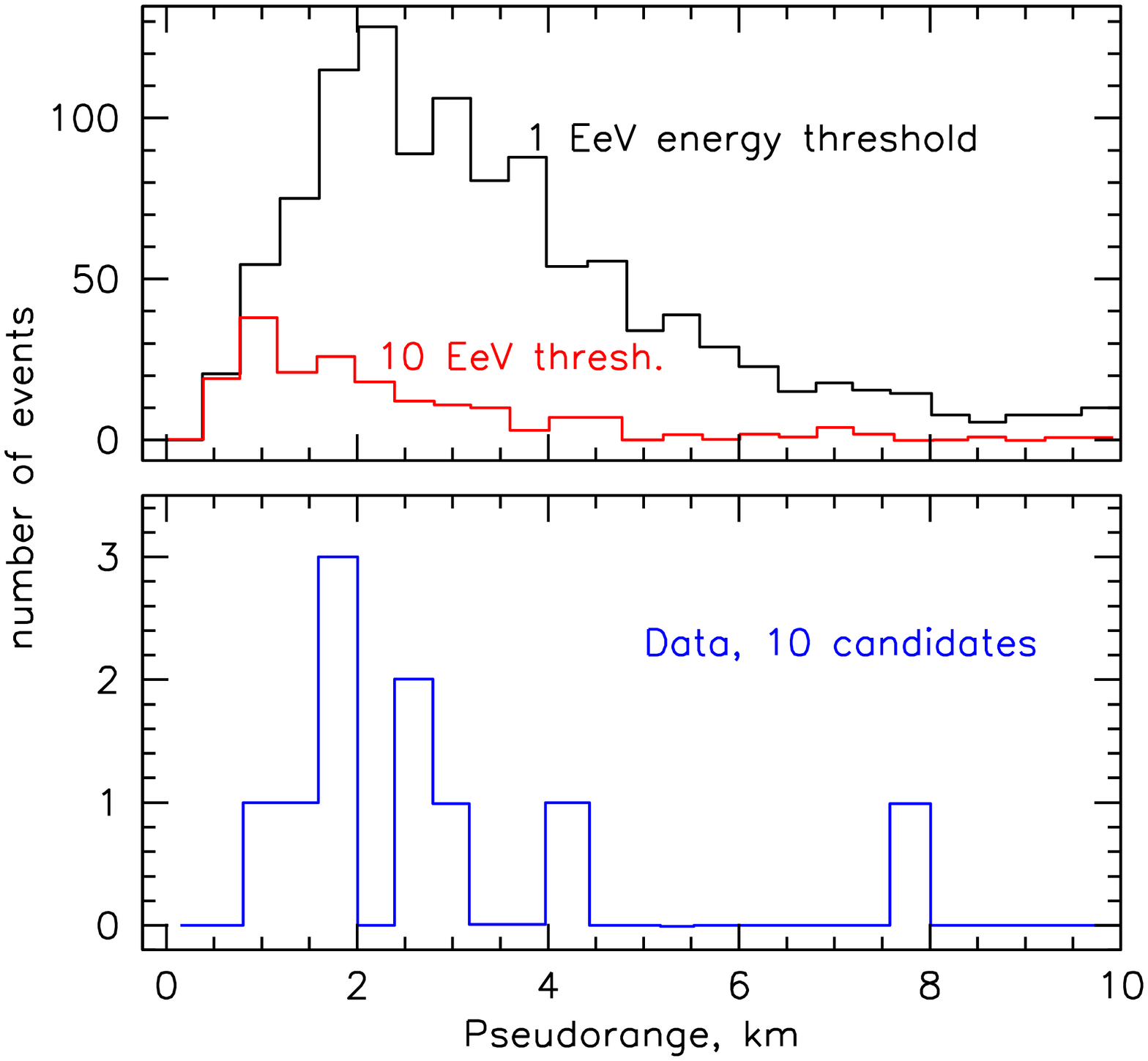}
  \caption{\it
Upper: Pseudorange distributions of simulated events for an AMBER
array with a 1 EeV energy threshold at 10~km distance (upper curve)
and a 10~EeV threshold at 10~km (lower curve). Lower: 
Pseudorange distribution of 10 candidate events measured in
recent AMBER data taken over several months.
  \label{amber_sims}}
\end{figure}

EAS observatories have also demonstrated capabilities for neutrino
observations, but will require substantial increases in their apertures
before such techniques can become practical in elucidating the
GZK neutrino spectrum. Neutrino-induced showers are also likely
to be highly-inclined relative to typical proton showers, and thus
become problematic for observation with ground arrays, which
suffer from more severe systematics at high zenith angles. In contrast
fluorescence methods (and possibly the MBR methods we describe here) 
can readily observe such showers, since the geometry is no less
favorable for horizontal than for vertical showers. Thus MBR observations
may help to greatly expand the neutrino apertures of air shower
observatories, by extending the duty cycle for "quasi-fluorescence"
observations, perhaps by an order of magnitude or more.
Figure~\ref{Ambergeom} gives a schematic view of how such
methodology relates to other implementations of ultra-high
enery cosmic ray air shower detection.

\subsection{The AMBER system.}

%{\bf The Radio Bremsstrahlung Impulse Detector (RaBID).}~~
Following the indications of stronger-than-expected emission from the two
accelerator experiments detailed above, we have moved ahead to 
develop a prototype of
a system
that could be used to search for detectable microwave emission
from actual air showers.  This system is built around a custom 
compact-PCI digitizer and data acquisition system, which we designate
the Radio Bremsstrahlung Impulse Detector (RaBID).
We have chosen the components and size of the prototype system
such that it can be duplicated at low cost with mostly commercial parts.
The proposed system, incorporating the RaBID prototype, is designated
the AMBER for Air-shower Microwave Bremsstrahlung Experimental Radiometer.
AMBER is currently operating 
on the rooftop of Watanabe Hall at the University of Hawaii at M\={a}noa
(UHM) in Honolulu, Hawaii, pictured in Fig.~\ref{fig:rabidpic1}.

In its current configuration, the AMBER unit consists of a
dual-band (C and Ku), dual-polarization feed horn array at the prime focus of a
1.8~m off-axis parabolic dish.  The array is in a diamond-shaped
configuration where each feed is $\sim5.2^\circ$ from its nearest
neighbor.  Each feed produces four channels of signal which are
amplified and down-converted in Low Noise Blocks (LNB) and then
conveyed to the RaBID DAQ via RG11 coaxial cable, as shown in
Figure~\ref{fig:rabid_scheme}.  The RaBID DAQ consists of a pair of
RaBID cards located inside a compact PCI (cPCI) crate, along with a
cPCI CPU for data collection and logging.  At the RaBID card input, the
down-converted LNB outputs are measured with RF power monitor (MAX4003)
chips, which provide  output proportional to the received RF
power, with approximately 70 ns integration time.  This power level
is sampled with a 32MSa/s ADC and processed inside a Field
Programmable Gate Array.  

These digitized samples are processed in 3
parallel paths: (1) all C-band samples are logged into a hardware
histogrammer, which allows optimum threshold-riding with varying
background (2) a trigger threshold is set based upon the histogram
values; and (3) a circular buffer holds the samples in time sequence to be
read out into the CPU upon detection of a trigger condition.  In order
to avoid biases in the triggering, each feed horn channel (of 12 total) is
triggered separately, at minimum possible threshold, and the trigger
times (corresponding to different transit times across the array field
of view) are analyzed in the stored data.  All sample times are
recorded with respect to a common clock, which is synchronized to GPS
via Network Time Protocol.  An external trigger port (not shown) is
available for forcing readout when observing in conjunction with another
detector.

\subsubsection{AMBER Results.}

Since initial commissioning of the AMBER system began in
mid-2005, we have accumulated about 8 months of data under
stable operating conditions, most of which has been analyzed to search for EAS-like
events. Because AMBER lacks a ground-truth EAS array to
validate any observed signals, any candidates that are
observed remain only putative at best. However, we may
test a sample of such candidates for similarity 
to expectations from our simulations, and we have done this 
for a large data sample taken through the spring of this 
year, with results that support the potential for
EAS measurements by an AMBER array.

Data analysis for AMBER events involves several steps
which significantly improve the signal-to-noise ratio of
the raw data. First, because the MBR signal is by nature
unpolarized, we can combine the recorded power in the
two independent polarizations, improving the SNR by a factor
of order $\sqrt{2}$. Also, the signal arrives over
many sequential 70~ns time bins, whereas a significant
fraction of the noise is
broadband and largely uncorrelated between successive time
bins. These statements are equivalent to saying that the
spectral bandwidth of the signal is much less than that of
the noise, and under such conditions we may apply 
Wiener filtering (also known as {\em optimal} filtering) to
remove the out-of-band noise component, and properly weight
the in-band noise components.

Examples of the effects of this analysis are shown for a
candidate event in Fig.~\ref{Amber_ev1}. On the left side
the raw event that triggered the system at C-band (4~GHz) is
shown, with the upper (earlier) signals from feed 4
and the later signals from feed 1 (lower left), indicating a downgoing
event. On the right-side panes, the signals are shown for the
combination of polarization co-adding and Wiener-filtering,
with a marked improvement in overall SNR and resulting timing.
Although the actual range to the event cannot be determined
directly, we calculate a pseudorange based on the assumption
that the feed-crossing signals were moving at the speed of light
over the known angle between feeds. These pseudorange values
can then be compared to simulations for actual EAS events.

Over the several months' observation period where the data has
the highest quality, we have selected a sample of candidate events
based on criteria derived from EAS expectations. Additional
examples of such candidates are shown in  Fig.~\ref{amber_evs}.
Here the projected elevations and azimuths for each feed are
shown in each event pair. We find that downgoing events
predominate in our current sample. This is expected from a true
EAS sample, but without an independent air-shower tag for any
given event, we cannot yet reject the possibility of anthropogenic
origin.

We can however compare the derived pseudorange distributions for
both simulations and actual data to determine if the 
candidates observed in our event sample are drawn from a
distribution that is consistent with what is expected from
actual EAS events. To do this we have developed a Monte Carlo
simulation code from which we can extract the pseudorange value
for events with various detectability thresholds and an
energy spectrum consistent with the known UHECR energy spectrum.

Results from this analysis are shown in Fig.~\ref{amber_sims}.
The upper pane shows the simulations for two energy threshold
at a distance of 10~km, and the lower pane shows the results for
a sample of current candidate events. While these candidates cannot
be proven to be EAS events without independent evidence from an
air shower array, they do appear at least consistent to first
order with the expectations from EAS events, and they demonstrate
that an AMBER array has the basic detector characteristics to
make measurements that are necessary
to establish MBR observations as a viable EAS detection methodology.

In summary, we have proceeded as far as possible with MBR studies
in the absence of coincident EAS ground-truth array measurements.
Efforts are now underway to deploy an AMBER testbed array 
within the Auger Observatory in Argentina. 

We thank the excellent support from staff at both Argonne
National Laboratory, and the Stanford Linear Accelerator
Center, especially the Experimental Facilities Division at
SLAC. This work was supported through the Department of Energy
High Energy Physics Division at the major laboratories,
through DOE Outstanding Junior Investigator Awards to P. Gorham
and D. Saltzberg, and through the DOE High Energy Physics Division
University Program.

\end{document}